\documentstyle[12pt]{article}
\def\s{\sigma}
\def\sq{\sqrt}
\def\p{\partial}
\def\d{\delta}

\begin{document}
\vspace{-2.0cm}
\bigskip

\begin{center}
{\Large \bf
Noncommutativity in open  string: New results in a gauge independent analysis}
\vskip .8 true cm
{\bf Rabin Banerjee ,} \footnote{rabin@bose.res.in}
{\bf Biswajit Chakraborty}\footnote{biswajit@bose.res.in}
\vskip .5cm
S. N. Bose National Centre for Basic Sciences \\
JD Block, Sector III, Salt Lake City, Calcutta -700 098, India.\\
\vskip .2cm
and\\
\vskip .2cm
{\bf Subir Ghosh}\footnote{sghosh@isical.ac.in}
\vskip .50 true cm
 Physics and Applied Mathematics Unit\\
Indian Statistical Institute \\
203 B.T.Road, Calcutta 700 035, India.\\
\end{center}
\bigskip

\centerline{\large \bf Abstract}
Noncommutativity in an open string moving in a background Neveu-Schwarz
field is investigated in a gauge independent Hamiltonian approach, 
leading to new results. The noncommutativity
is shown to be a direct consequence of the non-trivial boundary conditions, which,
contrary to several approaches, are not treated as constraints. We find that the
noncommutativity persists for all string points. In the conformal gauge our results reduce
to the usual noncommutativity at the boundaries only.\\

\newpage
\section{Introduction}
The study of open string, in the presence of a background Neveu-Schwarz two-form
field $B_{\mu\nu}$, leading to a noncommutative  structure
has recently evoked considerable interest \cite{sw,dn}. This structure manifests in the
noncommutativity in the spacetime coordinates of D-branes, where the end points of the string
are attached. Different approaches have been
adopted to obtain this result. A Hamiltonian operator treatment was
provided in \cite{chu} and a world sheet approach in \cite{som}. Also, an
alternative Hamiltonian (Dirac \cite{di}) approach based on regarding the Boundary Conditions (BC)
as constraints was given in \cite{ard}; the corresponding Lagrangian (symplectic)
version being done in \cite{bra}. The interpretation of the BC as
primary constraints usually led to an infinite tower of second class constraints
\cite{tez}, in contrast to the usual Dirac formulation of constrained
systems \cite{di, hrt}. Some other approaches to this problem have been discussed in \cite{zab, and}.
As has been stressed in \cite{sw}, it is very important to
understand this noncommutativity from different perspectives.

In the present work, we provide an exhaustive analysis of the noncommutativity in
open string theory moving in the presence of a constant Neveu-Schwarz field, in
the conventional Hamiltonian framework. In contrast to the usual studies, our model
of string theory is very general in the sense that no gauge is fixed at the
beginning. Let us recall that all computations of noncommutativity,
mentioned before, were done in the
conformal gauge. Our gauge independent analysis yields a new noncommutative structure, which correctly
reduces to the usual one in conformal gauge. This shows the compatibility of the
present analysis with the existing literature. In the general case, the noncommutativity is
manifested at all points of the string, in contrast to conformal gauge results
where it appears only at the boundaries. Indeed, in this gauge independent scheme,
one finds a noncommutative algebra among the coordinates, even for a free string, a fact that was
not observed before. Expectedly, this noncommutativity vanishes in the conformal gauge.
Note however, that there is no gauge for which noncommutativity vanishes in the interacting theory.
To gain further insight, both the Polyakov and Nambu-Goto (NG) formalisms of string theory
have been studied.

At the outset, let us point out the crucial difference between existing Hamiltonian analysis
\cite{ard} and our approach. This is precisely in the interpretation of the BC arising in the string
theory. The general consensus has been to consider the BCs as primary constraints of the
theory and attempt a conventional Dirac constraint analysis \cite{di}. The aim is to induce
the noncommutativity in the form of Dirac Brackets between coordinates. The subsequent analysis
turns out to be ambiguous since it involves the presence of $\delta (0)$-like factors,
(see Chu and Ho in \cite{ard}).
Different results are obtained depending on the interpretation of these factors.

We, on the other hand, do not treat the BCs as constraints, but show that they can be
systematically implemented by modifying the canonical Poisson Bracket (PB) structure. In this
sense our approach is quite similar in spirit to that of Hanson, Regge and 
Teitelboim \cite{hrt}, where modified PBs were
obtained for the free NG string, in the orthonormal gauge, which is the counterpart of
the conformal gauge in the free Polyakov string.

The paper is organized as follows: In section {\bf 2}, the gauge
independent analysis of free Polyakov string is discussed. This
also helps to fix the notations. The free NG string is developed
in section {\bf 3} for a comparison. A new structure, in the form
of an interpolating action is presented in section {\bf 4}, which
connects the Polyakov and NG actions in a smooth way. It also
highlights the role of the boundary conditions in the present
context. The noncommutativity is revealed in a gauge independent analysis, in
free Polyakov model in section {\bf 5}, which incidentally is a
new result. Section {\bf 6} discusses the noncommutativity in the interacting
theory in the Polyakov  formulation and section {\bf 7} does the
same in the NG formalism. The paper ends with a conclusion in
section {\bf 8}.

\section{The free Polyakov string}

In order to study the various ramifications of different formulations
of string theory, let us first consider the free Polyakov string action,
$$S_P= -{1\over 2} \int_{-\infty}^{+\infty} d\tau\int _0^\pi d\s {\sq {-g}}g^{ab}{\p}_aX^{\mu}{\p}_bX_{\mu}\eqno(1)$$
where $\tau $ and $\s $ are the usual world-sheet parameters and
$g_{ab}$, up to a Weyl factor, is the induced metric on the
world-sheet. $X^{\mu}(\xi )$ are the string coordinates in the
D-dimensional Minkowskian target space with metric $G_{\mu
\nu}=diag (-1,1,1..,1)$. This action has the usual Poincare, Weyl
and diffeomorphism invariances. Contrary to the usual approach of
working in the reduced space by choosing the conformal gauge at
the very beginning, we prefer to carry out the analysis in the
complete space by regarding both $X^{\mu}$ and $g_{ab}$ as
independent dynamical variables \cite{holt}. The canonical momenta
are,
$$\Pi_{\mu}={\d {\cal L}_P \over \d (\p_0X^{\mu})}=-{\sq {-g}}\p^0X_{\mu}$$
$$\pi_{ab}={\d {\cal L}_P \over \d (\p_0g^{ab})}=0 \eqno(2)$$
It is clear that while $\Pi_{\mu}$ is a genuine momenta, $\pi_{ab}\approx 0$
are the primary constraints of the theory. To determine the secondary constraints,
one can either follow the traditional Dirac's Hamiltonian approach, or just read
it off from the equation obtained by varying $g_{ab}$ since this is basically a
Lagrange multiplier. This imposes the vanishing of the symmetric energy-momentum
tensor,
$$T_{ab}={2\over {\sq {-g}}}{\d S_P\over \d g^{ab}}=-\p_aX^{\mu}\p_b{X_{\mu}}
+{1\over 2}g_{ab}g^{cd}\p_cX^{\mu}\p_dX_{\mu}=0 \eqno(3)$$
Because of the Weyl invariance, the energy-momentum tensor is traceless,
$${T^a}_a=g^{ab}T_{ab}=0 \eqno(4)$$
so that only two components of $T_{ab}$ are independent. These components, which
are the constraints of the theory, are given by,
$$\chi_1= gT^{00}=-T_{11}={1\over 2}(\Pi^2+(\p_1X)^2)=0$$
$$\chi_2= {\sq {-g}}{T^0}_1= {\Pi} .{\p}_1X =0 \eqno(5)$$
The canonical Hamiltonian obtained from (1) by a Legendre transformation is given by,
$$H=\int d\s {\sq {-g}}{T^0}_0= \int d\s {\sq {-g}}({1\over 2g_{11}}\chi_1+{g_{01}\over {\sq {-g}}g_{11}}\chi_2)\eqno(6)$$
Expectedly, the Hamiltonian turns out to be a linear combination of the constraints.

Just as variation of $g_{ab}$ yields the constraints, variation of $X^{\mu}$
gives the equation of motion,
$$\p_a({\sq {-g}}g^{ab}\p_bX^{\mu})=0 \eqno(7)$$
Finally, there is a mixed BC {\footnote {It is a mixed boundary
condition in the sense that ${\p}^1X^{\mu}=g^{11}\p_1X^{\mu}+g^{10}\p_0X^{\mu}$
will consist of both $\tau $ and $\s $ derivatives.}},
$$\p^1X^{\mu}(\tau ,\s)\vert_{\s=0,\pi}=0 \eqno(8a)$$
where the string parameters are in the region $-\infty \leq \tau \leq +\infty$,
$0\leq \s \leq \pi$. In the covariant form involving phase space variables, this
is given by
$$(\p_1X^{\mu}+{\sq {-g}}g^{01}\Pi^{\mu})\vert_{\s=0,\pi}= 0. \eqno(8b)$$
It is quite clear that the above boundary conditions are
incompatible with the first of the basic Poisson brackets (PB),
$$\{X^{\mu}(\tau ,\s ),\Pi_{\nu}(\tau ,\s')\} = \d^{\mu}_{\nu}\d (\s - \s')$$
$$\{g_{ab}(\tau ,\s ),\pi^{cd}(\tau ,\s')\}={1\over 2}({\d^c}_a{\d^d}_b+ {\d^d}_a{\d^c}_b)\d (\s -\s')\eqno(9)$$
where $\d (\s -\s')$ is the usual one-dimensional Dirac delta function. We would
also like to mention that there is an apparent contradiction of the
constraint $\pi_{ab}\approx 0$ with the PB (9). However this equality is valid in
Dirac's ``weak" sense only, so that it can be set equal to zero only after the relevant
brackets have been computed. These weak equalities will be designated by $\approx$,
rather than an equality, which is reserved only for a strong equality. In this
sense, therefore, there is no clash between this constraint and the relevant PB.
Indeed, we can even ignore the canonical pair $(g_{ab}, \pi^{cd})$. From the
basic PB, it is easy to generate a first class (involutive) algebra,
$$
 \{\chi _1(\sigma ),\chi _1(\sigma ')\}=4(\chi _2(\sigma )+\chi _2(\sigma '))\partial _{\s}\delta(\sigma -\sigma '),$$
$$
\{\chi _2(\sigma ),\chi _1(\sigma ')\}=(\chi _1(\sigma )+\chi
_1(\sigma '))\partial _{\s}\delta(\sigma -\sigma '), $$
$$
 \{\chi
_2(\sigma ),\chi _2(\sigma ')\}=(\chi _2(\sigma )+\chi
 _2(\sigma '))\partial _{\s }\delta(\sigma -\sigma ').
\eqno(10)$$
The situation is quite similar to usual electrodynamics. There the Lagrange multiplier is $A_0$,
which corresponds to $g_{ab}$ in the string theory. The multiplier $A_0$ enforces the
Gauss constraint just as $g_{ab}$ enforces the constraints $\chi_1$ and $\chi_2$. Furthermore, the
 Gauss constraint generates the time independent gauge transformations, while
$\chi_1$, $\chi_2$ generate the diffeomorphism transformations.

The BC (8), on the other hand, is not a constraint in the Dirac sense \cite{di}, since it
is applicable only at the boundary {\footnote {We are therefore differing from recent
approaches \cite{ard} which regard the BCs as Dirac constraint. Our views are similar to
those of \cite{hrt}, who discuss the free NG string.}}. Thus, there has to be an appropriate
modification in the PB, to incorporate this condition. This is not unexpected and
occurs, for instance, in the example of a free scalar field $\phi (x) $ in $1+1$ dimension,
subjected to periodic BC of period, say, $2\pi$ ($\phi (t,x+2 \pi )= \phi (t,x)$). There
the PB between the field $\phi (t,x)$ and its conjugate momentum $\pi (t,x)$ are
given by,
$$\{\phi (t,x),\pi (t,y)\}= \d_P(x-y) \eqno(11a)$$
where,
$$\d_P(x-y)={1\over {2\pi}}\sum_{n\in {\cal Z}}e^{in(x-y)} \eqno(11b)$$
is the periodic delta function of period $2\pi$ and occurs in the closure properties
of the basis functions $e^{inx}$ for the space of square integrable functions, defined
on the unit circle $S^1$. In fact, one can easily show that this PB  algebra is obtained
automatically if one starts with the canonical harmonic oscillator algebra for
each mode in the Fourier space.

Before actually computing the modifications in the usual PB, let us take a look
at the free NG action.

\section{The free Nambu-Goto action}

The NG action is given by,
$$S_{NG}=-\int d\tau d\s [({\dot X}.X')^2-{\dot X}^2{X'}^2]^{1\over 2} \eqno(12)$$
where ${X'}^{\mu}={\p X^{\mu}\over \p \s}=\p_1X^{\mu}$ and ${\dot X}^{\mu}={\p X^{\mu}\over \p \tau}=\p_0X^{\mu}$
have been introduced for notational convenience. Note that, here the induced
metric on the world-sheet has not been introduced, as we are exclusively working
with $\tau ,\s $ variables. A systematic constrained analysis of this action
has already been carried out in \cite{hrt} and here we just give the results.
This will also put the analysis of the Polyakov string formulation in a proper perspective.
The Euler equations are,
$$\p_0\Pi^{\mu}+\p_1K^{\mu}=0 \eqno(13a)$$
where,
$$\Pi_{\mu}={\p L_{NG}\over \p {\dot X}^{\mu}}={(X'.\dot X)X'_{\mu}-X'^2\dot X_{\mu})\over {[(X'.\dot X)^2-X'^2\dot X^2]^{1\over 2}}}$$
and $$K_{\mu}={\p L_{NG}\over \p X'^{\mu}}={(X'.\dot X)\dot X_{\mu}-\dot X^2X'_{\mu})\over {[(X'.\dot X)^2-X'^2\dot X^2]^{1\over 2}}}
\eqno(13b)$$
The definition of the momenta  $\Pi _\mu $ immediately leads to two primary constraints,
$$\Pi^2+X'^2 \approx 0 \eqno(14a)$$
$$\Pi .X' \approx 0 \eqno(14b)$$
And the BCs are,
$$K_{\mu}(\tau ,0)=K_{\mu}(\tau ,\pi)=0 \eqno(15)$$
A simple comparison shows that although the constraints in the Polyakov (5) and NG formulations (14)
have the same functional form, the BCs do not share this property (8, 15).

If one wants to match the BCs also, it is necessary to choose a particular gauge. In
the NG formulation one can take the orthonormal gauge conditions \cite{hrt},
$$\lambda_{\mu}(X^{\mu}(\tau ,\s)-{\tau \over \pi}{\cal P}^\mu )\approx 0 ,$$
$$\lambda_{\mu}(\Pi^{\mu}(\tau ,\s)-{1 \over \pi}{\cal P}^\mu )\approx 0 ,\eqno(16)$$
where $\lambda_{\mu}$ is an arbitrary constant D-vector and
 ${\cal P}^\mu =\int^{\pi}_0 d\s \Pi^\mu $ denotes the conserved momentum,
following from the equations of
motion.

 With these conditions the NG action
{\it weakly} (i.e. on the constraint surface) reduces to,
$$S_{NG}\approx {1\over 2}\int d\tau d\s ({\dot X}^2-{X'}^2)\eqno(17)$$
while the BCs become the usual Neumann type:
$${X'}^{\mu}\vert_{\s =0,\pi}\approx 0. \eqno(18)$$
The orthonormal gauge corresponds to the conformal gauge in the Polyakov
formulation, so that the induced metric $g_{ab}=\eta_{ab}=diag(-1,1)$. Then the Polyakov
action (1) and the BC (8) exactly match with the corresponding expressions for the NG case.

\section{The interpolating free string action}

From our analysis in the previous sections, we saw that the NG and Polyakov actions,
along with their BCs, agreed in the orthonormal and conformal gauge respectively. Here we
discuss a new form of the action that interpolates between the two forms, without the
need of any gauge fixing.

The starting point is to rewrite the free NG action in a first order form \cite{holt},
incorporating
the constraints,
$${\cal L}_I= \Pi_{\mu}{\dot X}^{\mu}-{\cal H}= \Pi_{\mu}{\dot X}^{\mu}+{1\over 2}{\lambda}
({\Pi_{\mu}}^2+X_{\mu}'^2)+\rho \Pi_{\mu}X'^{\mu} \eqno(19)$$ Note
that there is no contribution from the canonical Hamiltonian,
obtained by a Legendre transformation, as it vanishes identically-
a typical feature of a reparametrisation invariant theory like
this. So the expression of the Hamiltonian $\cal H$ appearing here
is just a linear combination of the constraints, with $\lambda$
and $\rho$ playing the roles of Lagrange multipliers enforcing the
respective constraints. Consequently, the time evolution of the
system here is given by a gauge transformation.

Coming back to (19), we observe that $\Pi_{\mu}$ appears here as an auxiliary variable.
It is thus possible to eliminate it using its equation of motion. We find,
$${\cal L}_I=-{1\over 2\lambda}({\dot X}^2_{\mu}+ 2\rho {\dot X}_{\mu}{X'}^{\mu}+({\rho}^2-{\lambda}^2){X'}_{\mu}^2)\eqno(20)$$
This is the cherished form of our interpolating Lagrangian.

If $\rho$ and $\lambda$ are eliminated by their respective equations of motion,
$$\rho =-{{\dot X}_{\mu}X'^{\mu}\over {X_{\mu}'}^2}$$
$$\lambda^2=-{h\over {X'}_{\mu}^2{X'}_{\nu}^2} \eqno(21)$$
then the above Lagrangian (20) reduces to the NG form (12).

If, on the other hand, we identify $\rho$ and $\lambda$ with the following contravariant
components of the world-sheet metric,
$$g^{ab}=(-g)^{-{1\over 2}}\pmatrix{{1\over \lambda} & {\rho \over \lambda}\cr {\rho \over \lambda} &
{(\rho^2-\lambda^2)\over {\lambda}}}\eqno(22)$$
then the action reduces to the Polyakov form (1). In this sense, therefore, the Lagrangian in (20)
is referred to as an interpolating Lagrangian \cite{mk}. Also, note that with this mapping, the Hamiltonian
read-off from (19) just reproduces the result (6).

Next, the BC is analysed. In general the BC of an open string is given by,
$$K^{\mu}={\p L\over \p X'_{\mu}}\vert_{\s =0,\pi}=0.$$
From the interpolating Lagrangian (20), we find,
$$K^{\mu}=({\rho \over \lambda}{\dot X}^{\mu}+{\rho^2- \lambda^2 \over \lambda}X'^{\mu})\vert_{\s=0,\pi}=0 \eqno(23)$$
at $\s =0,\pi$. Now using the expressions (21) for $\rho$ and $\lambda$, we recover the
usual BC (15) for NG string.

To get the BC for Polyakov string, it is useful to rewrite (23) in terms of phase space
variables, $X^{\mu}$ and $\Pi_{\mu}$, as
$$K^{\mu}=(\rho \Pi^{\mu}+ \lambda {X'}^{\mu})\vert_{\s =0,\pi}=0 \eqno(24)$$
where
$$\Pi^{\mu}={\p {\cal L}_I \over \p {\dot X}_{\mu}}=-{1 \over \lambda}({\dot X}^{\mu}+ \rho {X'}^{\mu}) \eqno(25)$$
Now identifying $\rho$ and $\lambda $ with the metric components, it is easy to check that the Polyakov
form of BC (8) is reproduced. Hence it is possible to interpret either of (23) or (24) as an interpolating
BC.

It is noteworthy that although the Polyakov BC can be expressed in terms of pure phase space
variables, the Nambu-Goto BC cannot be done so, because of the presence of velocities in $\rho$ (see (21)). This
is an important distinction when it comes to the study of the modification in the basic algebra,
as will become evident in the next section.

\section{Boundary Conditions and modified brackets for a free theory}

Before discussing the mixed type condition, that emerged in a completely gauge independent
formulation of the Polyakov action, consider the simpler Neumann type condition (8) that leads
to $(\p_1X^{\mu})\vert_{\s =0,\pi}=0$ in an orthonormal (conformal) gauge.

Since the string coordinates $X^{\mu}(\tau ,\s )$ transform as a
world-sheet scalar under its reparametrisation, it will be even
more convenient to get back to our scalar field $\phi (t,x)$
defined on $1+1$ dimensional space-time, but with the periodic BC
of $2\pi $ replaced by Neumann BC
$$\p_x\phi \vert_{\s =0,\pi}=0 \eqno(26)$$
at the end points of a 1-dimensional box of compact size, i.e. of length $\pi$.
Correspondingly, the $\d_P(x)$ appearing there in the PB (11)-consistent with periodic BC-
have to be replaced now with a suitable ``delta function" incorporating Neumann BC, rather
than periodic BC. Interestingly, such a ``delta function" is not difficult to construct from
purely algebraic arguments.

One starts by noting that the usual properties of a delta function is also satisfied by $\d_P(x)$:
$$\int_{-\pi}^{+\pi}dx'\d_P(x'-x)f(x')=f(x) \eqno(27)$$
for any periodic function $f(x)=f(x+2\pi )$ defined in the interval $[-\pi, +\pi ]$. Let us now restrict to the case of even (odd)
functions $f_\pm (-x)=\pm f_\pm (x)$. Then it can be easily seen that the above integral (27) reduces to,
$$\int_0^{\pi}dx' \Delta_{\pm}(x',x)f_\pm (x')=f_\pm (x) \eqno(28)$$
where,
$$\Delta_{\pm}(x',x)= \d_P(x'-x)\pm \d_P(x'+x) \eqno(29a)$$
Using (11b), the explicit form of $\Delta_{+}(\s',\s )$, in particular, can be given as,
$$\Delta_{+}(\s ,\s' )={1\over \pi }+{1\over \pi }\sum_{n \neq 0}cos(n\s' )cos(n\s ) \eqno(29b)$$
We will not have to deal with $\Delta_{-}(\s' ,\s )$ henceforth in our paper, for reasons
explained below.

Since any function $\phi (x)$ defined in the interval $[0,\pi ]$ can be regarded as a part of an
even/odd function $f_\pm (x)$ defined in the interval $[-\pi ,\pi ]$, both $\Delta_{\pm}(\s',\s )$
act as delta functions defined in half of the interval at the right i.e.$[0,\pi ]$ (28).
It is still not clear which of these $\Delta (x',x)$ functions should replace
$\d_P(x'-x)$ in the PB relation. We can invoke the Neumann BC, at this stage, to fix
the matter. To see this, consider the Fourier decomposition of an arbitrary function
$f(x)$ satisfying periodic BC, ($f(x)=f(x+2\pi )$)
$$f(x)=\sum_{n\in {\cal Z}}f_ne^{inx}. \eqno(30)$$
Clearly,
$$f'(0)= i\sum_{n>0}n(f_n-f_{-n})$$
$$f'(\pi )=i\sum_{n>0}(-1)^nn(f_n-f_{-n}) \eqno(31)$$
Now for even(odd) functions, the Fourier coefficients are related as,
$$f_{-n}= \pm f_n \eqno(32)$$
so that Neumann's BC
$$f'(0)=f'(\pi )=0 \eqno(33)$$
are satisfied if and only if $f(x)$ is even. Therefore, one has to regard the scalar field $\phi (x)$ defined
in the interval $[0, \pi ]$ and subjected to  Neumann BC (26) as a part of an even periodic function
$f_+ (x)$ defined in the  extended interval $[-\pi, +\pi ]$. It thus follows that 
the appropriate PB for the
scalar theory is given by,
$$\{\phi (t,x), \pi (t,x')\}= \Delta_{+} (\s ,\s')$$
It is clearly consistent with Neumann BC as $\p_{\s }\Delta_{+}(\s ,\s')\vert_{\s =0,\pi }=\p_{\s'}\Delta_{+}(\s ,\s')\vert_{\s =0,\pi } =0$
is automatically satisfied.
It is straightforward to generalise it to the string case, where it is given by,
$$\{X^{\mu}(\tau ,\s ), \Pi_{\nu }(\tau, \s') \}= \d^{\mu}_{\nu}\Delta_{+}(\s ,\s')\eqno(34a)$$
and the Lorentz indices are playing the role of ``isospin" indices, as viewed from the world-sheet.
This form first appeared in \cite{hrt}.
Observe also that the other brackets
$$\{X^{\mu}(\tau ,\s ),X^{\nu }(\tau ,\s')\}=0 \eqno(34b)$$
and
$$\{\Pi^{\mu}(\tau ,\s ), \Pi^{\nu }(\tau ,\s')\}=0 \eqno(34c)$$
are consistent with the BCs and hence remain unchanged.

For a gauge independent analysis, the Nambu-Goto BC poses problems since it cannot be
expressed in phase space variables.
To overcome this problem it is necessary to fix a gauge and this was elaborated in section {\bf 3}. The generalisation of this in the interacting NG string will be given later in section {\bf 7}. Here we take recourse to the mixed condition
(8) that occurs in the Polyakov string. A simple inspection
shows that this is also compatible with the modified brackets (34a, 34c), but not with
(34b). Hence the bracket among the coordinates should be altered suitably. We therefore make an ansatz,
$$\{X^{\mu}(\tau ,\s ),X^{\nu }(\tau ,\s')\}=C^{\mu \nu}(\s ,\s') \eqno(35a)$$
where,
$$C^{\mu \nu}(\s ,\s')=-C^{\nu \mu}(\s' ,\s). \eqno(35b)$$
Imposing the BC (8) on this algebra, we get,
$$\p_{\s'}C^{\mu\nu}(\s ,\s')\vert_{\s'=0,\pi}=\p_{\s }C^{\mu\nu}(\s ,\s')\vert_{\s =0,\pi}
=-{\sq {-g}}g^{01}\{\Pi^{\mu}(\tau ,\s),X^{\nu}(\tau,\s')\}$$
$$={\sq {-g}}g^{01}G^{\mu \nu}\Delta_{+}(\s ,\s') \eqno(36)$$
For an arbitrary form of the metric tensor, it might be technically problematic to find
a solution for $C^{\mu\nu}(\s ,\s')$. However, for a restricted class of metric {\footnote {Such
conditions also follow from a standard treatment of the light-cone gauge \cite{pol}}} that
satisfy,
$$\p_1g_{ab}=0 \eqno(37)$$
it is possible to give a quick solution of $C^{\mu\nu}(\s ,\s')$ as,
$$C^{\mu\nu}(\s ,\s')={\sq {-g}}g^{01}G^{\mu\nu}[\Theta(\s ,\s')-\Theta(\s',\s )] \eqno(38)$$
where the generalised step function $\Theta (\s ,\s')$ satisfies,
$$\p_{\s }\Theta (\s ,\s')= \Delta_{+}(\s ,\s') \eqno(39)$$
An explicit form of $\Theta $ is given by \cite{hrt},
$$\Theta (\s ,\s')={\s \over \pi}+{1\over \pi }\sum_{n\neq 0}{1\over n}sin(n\s )cos(n\s' )~,\eqno(40a)$$
having the properties,
$$\Theta (\s ,\s')=1~~~ for ~~\s >\s'~,$$
and$$\Theta (\s ,\s')=0 ~~~for~~ \s <\s'.\eqno(40b)$$
Using these relations, the simplified structure of noncommutative algebra follows,
$$\{X^\mu (\tau,\s ),X^{\nu}(\tau, \s' )\}=0~~~for ~~\s =\s'~~$$
$$\{X^\mu (\tau,\s ),X^{\nu}(\tau, \s' )\}=\pm {\sq {-g}}g^{01}G^{\mu \nu}~~~for ~~\s >\s'~~ and ~~\s < \s' \eqno(41)$$
respectively.
Thus a noncommutative algebra for distinct coordinates $\s \neq \s' $ of the string emerges
automatically in a free string theory if a gauge independent analysis is carried
out like this. But this non-commutativity can be made to vanish in gauges like conformal gauge, where $g^{01}=0$, thereby
restoring the usual commutative structure. However, the non-commutativity among the string coordinates {\it cannot}
be made to vanish in any gauge if the string is coupled to a constant external B-field,
as we show in the next section.

Before we conclude this section, we would like to mention that the essential structure of the involutive
algebra (10) is still preserved, only that $\delta (\s -\s')$ has to be replaced by $\Delta_+(\s ,\s')$.
And this is despite the fact that the original basic brackets (9) have now been modified to (34a,34c,41).
Indeed, using these relations, one can show that

$$  \{\chi _1(\s ),\chi _1(\s')\}=4(\chi _2(\s )+\chi _2(\s'))\partial _{\s }\Delta_{+}(\s ,\s'),$$
$$\{\chi _2(\s ),\chi _1(\s')\} =(\chi _1(\s )+\chi_1(\s'))\partial _{\s }\Delta_{+}(\s ,\s') \eqno(42) $$
$$ \{\chi_2(\s ),\chi _2(\s' )\}=(\chi _2(\s )+\chi_2(\s '))\partial _{\s }\Delta_{+}(\s ,\s' ).$$

Note that, contrary to the usual case, the right hand side vanishes identically on the
boundary. A crucial intermediate step in this derivation is to use the relation,
$$\{X'^{\mu}(\s ), X'^{\nu}(\s' )\}=0$$
which follows from the basic bracket (41).

\section{The interacting theory:  Polyakov formulation}

The Polyakov action for a bosonic string moving in the presence of a constant
background Neveu-Schwarz two-form field $B_{\mu \nu}$ is given by,
$$S_P= \int d\tau d\sigma (-{1\over 2} {\sq {-g}}g^{ab}{\p}_aX^{\mu}{\p}_bX_{\mu}
+e\epsilon^{ab}B_{\mu \nu}\p_aX^{\mu}\p_bX^{\nu})\eqno(43)$$
where we have introduced a `coupling constant' $e$ and $\epsilon^{01}=-\epsilon^{10}=+1$.
Here too we shall carry out a gauge independent analysis at the beginning, rather
than making use of any gauge fixing condition (like conformal gauge) right at this stage.
A usual canonical analysis leads to the following set of primary first class constraints,
$$gT^{00}={1\over 2}[(\Pi_{\mu}+eB_{\mu \nu}\p_1X^{\nu})^2+(\p_1X)^2]\approx 0 \eqno(44)$$
$${\sq {-g}}{T^0}_1= {\Pi} .{\p}_1X \approx 0 \eqno(45)$$
where
$$\Pi_{\mu}=-{\sq {-g}}\p^0X_{\mu}+eB_{\mu \nu}\p_1X^{\nu} \eqno(46)$$
is the momentum conjugate to $X^{\mu}$.

Likewise, the BC is given by,
$$(\p^1X^{\mu}+{1\over {\sq {-g}}}eB^{\mu \nu}\p_0X_{\nu})\vert_{\s =0,\pi }=0 \eqno(47)$$
Using phase space variables, this can be written in a completely covariant form as,
$$((\p_1X_{\rho}){M^{\rho}}_{\mu}+{\Pi}^{\nu}N_{\nu \mu})\vert_{\s =0,\pi } =0 \eqno(48)$$
where,
$${M^{\rho}}_{\mu}={1\over g_{11}}[{\d^{\rho}}_{\mu}-{2e\over {\sq {-g}}}g_{01}{B^{\rho}}_{\mu}
+e^2B^{\rho \nu}B_{\nu \mu}] \eqno(49a)$$
$$N_{\nu \mu}=-{g^{01}\over g^{00}{\sq {-g}}}G_{\nu \mu}-{1\over g_{11}}eB_{\nu \mu} \eqno(49b)$$
are two matrices.
This nontrivial BC leads to a modification in the original (naive) canonical PBs.

 Now
the BC (48) is recast as,
$$(\p_1X_{\mu}+\Pi^{\rho}(NM^{-1})_{\rho \mu})\vert_{\s =0,\pi } =0 \eqno(50)$$
The $\{X^\mu (\s), \Pi _\nu (\s ')\}$ PB is the same as that of the free string (34a).
Considering the general structure (35), we obtain,
$$\{\p_{\s }X^{\mu}(\s ),X^{\nu }(\s' )\}= \p_{\s }C^{\mu \nu}(\s ,\s') \eqno(51)$$
Putting the BC and exploiting (34a), we get
$$\partial _\s C_{\mu\nu}(\s ,\s ')\mid _{\s =0,\pi}=(NM^{-1})_{\nu \mu}\Delta _+(\s, \s')
\mid_{\s =0,\pi}. \eqno(52)$$
As we did in the free case, we restrict to the class of metrics defined by (37). Taking a cue
from the free theory, the solution for $C_{\mu\nu}(\s ,\s ')$ must involve the generalised
$\Theta $ function, introduced in (40). Splitting $(NM^{-1})_{\nu \mu}$ into its symmetric
$(NM^{-1})_{(\nu \mu)}$ and antisymmetric $(NM^{-1})_{[\nu \mu ]}$ components, a general
solution for $C_{\mu\nu}$ is given by,
$$
C_{\mu\nu}(\s ,\s ')=
{1\over 2}(NM^{-1})_{(\nu \mu )}[\Theta (\s, \s')-\Theta (\s', \s )]
+{1\over 2}(NM^{-1})_{[\nu \mu]}[\Theta (\s, \s')+\Theta (\s', \s )-1].\eqno(53)$$
Observe that, by demanding (35b), $(NM^{-1})_{(\nu \mu )}$ must be multiplied by an
antisymmetric combination of $\Theta$'s, which is precisely $[\Theta (\s ,\s')-\Theta (\s',\s )] $
Likewise, the other factor $(NM^{-1})_{[\nu \mu ]}$ must be multiplied by a symmetric
combination  $[\Theta (\s ,\s')+\Theta (\s',\s )] $, plus an undetermined constant. We
fix this constant to $(-1)$ by requiring that the vanishing result (41) in the free
case is retained for all $\s =\s' $ away from the boundary (using $\Theta (\s ,\s )={1\over 2}$).
An advantage of this normalisation is that by passing to the conformal gauge, where $g=-1$
and $g^{01}=0$, one obtains,
$$C_{\mu \nu}(\s ,\s')={\tilde B}_{\mu \nu}[\Theta (\s, \s')+\Theta (\s', \s )-1] \eqno(54a)$$
where,
$${\tilde B}_{\mu \nu}=-e[B(1+e^2B^2)^{-1}]_{\mu \nu} \eqno(54b)$$
which reproduces the standard non-commutative algebra in the presence of a background field
\cite{sw,dn,chu,som,ard,bra}.

It is evident that the modified algebra is gauge dependent, depending on the choice of
the metric. However, there is no choice, for which the non-commutativity vanishes. To
show this, note that the origin of the non-commutativity is the presence of non-vanishing
$\Pi^{\mu}$ term in the BC (48). If this can be eliminated, then the usual commutative algebra
is obtained. This requires $N_{\nu\mu }=0$. From (49b) this implies $B_{\mu\nu}$ and
$G_{\mu\nu}$ have to be proportional which obviously cannot happen, as the former is
an antisymmetric and the latter is a symmetric tensor. Hence non-commutativity
will persist for any choice of world-sheet metric $g_{ab}$. Specially interesting are the
expressions for noncommutativity (53) at the boundaries,
$$C_{\mu\nu}(0 ,0)=-C_{\mu\nu}(\pi ,\pi )={1\over 2}(NM^{-1})_{[\nu \mu ]}
~~,~~C_{\mu\nu}(0 ,\pi)=
-C_{\mu\nu}(\pi ,0)=
-{1\over 2}(NM^{-1})_{(\nu\mu)}~.\eqno(55)$$
It should be pointed out that in the conformal gauge, $(NM^{-1})$ does not have a symmetric component, so that
$$C_{\mu \nu }(0,\pi )=C_{\mu \nu }(\pi ,0)=0. $$

\section{The interacting theory: Nambu-Goto formulation }
Although the Polyakov and NG formulations for free strings are
regarded to be classically equivalent, there are some subtle
issues. Indeed the structures of BC's in the two formulations are
different as was also illuminated by our interpolating action.
Also more complications are expected in the presence of
interactions. Since the occurrence of noncommutativity is directly connected
with the BC's, it is therefore useful to study this feature in the
NG formulation. This motivates us to carry out an exhaustive
analysis of the classical relativistic string interacting with a
constant, second rank, antisymmetric tensor $B_{\mu\nu }$ in the
NG formulation in this subsection. Here we present a
generalisation of the analysis of Hanson, Regge and Teitelboim
\cite{hrt} for the free string, to show that the noncommutativity
appears directly from taking proper account of the boundary
conditions. The analysis for the free theory \cite{hrt} has
already been reproduced briefly in section {\bf 3}.

We start with the action,
$$S= \int_{-\infty}^{+\infty}d\tau \int ^\pi _0d\s [{\cal L}_0+eB_{\mu\nu }{\dot X}^\mu X'^\nu ]\eqno(56)$$
Here ${\cal L}_0$ denotes the free string Lagrangian density appearing in (12).
From the variation of the action, we obtain the following  equations of motion and the BC's,
$$\dot \Pi^\mu +K'^\mu =0, \label{eq} \eqno(57)$$
$$K ^\mu\vert_{\s=0,\pi} =0 ,\eqno(58)$$
where
$$
\Pi^\mu = {\p {\cal L} \over {\p {\dot X}_\mu }}
={\cal L}_0^{-1}(-X'^2\dot X^\mu +(\dot X.X')X'^\mu
)+eB^{\mu\nu } X'_\nu , \eqno(59)$$
$$
K^\mu = {{\p {\cal L}}\over {\p X'_\mu }}
={\cal L}_0^{-1}(-{\dot X}^2 {X'}^\mu +(\dot X.X')\dot X^\mu
)-eB^{\mu\nu } \dot X_\nu .\eqno(60)$$
The Primary constraints of the theory are,
$$\chi _1=(\Pi^\mu -eB^{\mu\nu }X'_\nu )^2 +X'^2\approx 0~~,~~\chi_2=\Pi.X'\approx 0 ~~.\eqno(61)$$
which are similar to those obtained in the Polyakov version (see (44,45)). Using the
standard canonical PB, it is straightforward to verify the diffeomorphism algebra (10).

A gauge independent analysis, as was done for the Polyakov formulation, is not
feasible here, since the BC involves time derivatives that are not eliminatable in terms
of the momenta. To properly account for the BCs, a gauge choice becomes necessary.
This is equally valid for a free theory. As was shown in \cite{hrt} and
discussed in section {\bf 3}, the free
theory becomes most tractable in the orthonormal gauge. Inspired by their choice,
we consider the following gauge conditions,
$$
\lambda _\mu (X^\mu -{{\cal P}^{\mu}\tau \over {\pi}})\approx
0~~,~~\lambda _\mu (\Pi^\mu -{{\cal P}^\mu \over \pi })\approx 0,
\eqno(62)
$$
which for $e=0$ reduce to the orthonormal gauge in free theory. Here $\lambda _\mu $ is a
constant D-vector. For our present analysis there arises no need to fix $\lambda _\mu $.
Same notations as in section {\bf 3} are used here.

Let us study the consequences of the gauge choice. From  the gauge conditions
(62), we obtain $\p_0(\lambda.\Pi)={\p_0(\lambda .{\cal P})\over \pi}=0 $ and together with
the equations of motion (57) this leads to $\p_1(\lambda
.K )=0$. Compatibility with the BC (58) then ensures that
$\lambda .K =0$ for all
$\sigma $. Again, from the gauge choice (62), we find $\lambda .X'=0$ and
$\lambda .\dot X=(\lambda
.\Pi)$ . In short, the following three exact relations are valid,
$$
\lambda .K =0~~,~~\lambda .X'=0~~,~~\lambda .\dot X=(\lambda
.\Pi).
\eqno(63)
$$
 From the defining equations (59),(60) and (62), we find,
$$
(\lambda .\Pi)=-{\cal L}_0^{-1}X'^2(\lambda .\dot X) +eB^{\mu\nu
}\lambda _\mu X'_\nu ,
\eqno(64)
$$
$$
\lambda .K = {\cal L}_0^{-1}(\dot
X.X')(\lambda .\dot X )-eB^{\mu\nu }\lambda _\mu \dot X_\nu =0.
\eqno(65)
$$
Using (62) once again we obtain,
$$
{\cal L}_0^{-1}X'^2=-1+e{\cal A} ~~,~~{\cal L}_0^{-1}({\dot X}.X')=e{\cal B} \eqno(66a)
$$
where,
$${\cal A}=\frac{B^{\mu\nu}\lambda _\mu X'_\nu}{\lambda . \Pi}~~,~~
{\cal B}=\frac{B^{\mu\nu}\lambda _\mu 
{\dot X}_\nu}{\lambda . \Pi}. \eqno(66b)$$
From now on we will work in the lowest nontrivial order in the coupling $e$.
The explicit expressions for ${\cal A}$ and ${\cal B}$ are not needed for the $O(e)$ results presented here. Recalling the explicit
form of ${\cal L}_0$ from (56), we find,
$${\cal L}_0\approx -(-\dot X^2X'^2)^{\frac{1}{2}}. \eqno(67) $$ Using (66) we obtain,
$$X'^2=-\dot X^2(1-2e{\cal A})~\rightarrow \dot X^2+X'^2\approx 2e{\cal A} \dot X^2 .$$
The $O(e)$ correction to the orthonormality vanishes in the free
theory ($e=0$), where $\dot
X.X'=0$, and $\dot X^2+X'^2=0 $ \cite{hrt}.
Now ${\cal L}_0 $ is simplified to,
$${\cal L}_0 \approx -(-\dot X^2X'^2)^{\frac{1}{2}}
\approx \dot X^2(1-e{\cal A})$$
$$\approx [\frac{1}{2}(\dot X^2-X'^2)-e{\cal A}(\dot X^2+X'^2)]
\approx \frac{1}{2}(\dot X^2-X'^2)\eqno(68).$$
Finally we recover the Lagrangian of the string coupled to $B_{\mu\nu }$, in this
particular gauge, to lowest order in the coupling $e$ as,
$$
{\cal L}= {1\over 2}(\dot X^2- X')+eB_{\mu\nu }\dot X^\mu X'^\nu
+O(e^2) \eqno(69)
$$
The equation of motion in this gauge is that of a free theory,
$$
(\partial _0^2- \partial _1^2)X^\mu=0,
\eqno(70) $$
but crucial modifications have appeared in the BC,
$$
(X'^\mu +{e\over N} B_{\mu\nu }\dot X^\nu )\vert_{\s =0,\pi}=0.
\eqno(71)
$$
In fact, the interaction have changed the BC from Neumann type in free theory to
a mixed one. Elimination of $\dot X_\mu $ from (58) reproduces the BC in
phase space,
$$
X'^\mu+e(M^{-1}B)^{\mu\nu}\Pi_\nu =0,
\eqno(72)
$$
 where $M^{\mu\lambda}=G^{\mu\lambda }-{{e^2}\over {N^2}}B^{\mu\nu}B_\nu
^\lambda $. It is amusing to note that this BC is identical to the one used
 in the Polyakov model in the conformal gauge  \cite{chu,ard} but in our case we should consider  $M^{\mu\lambda}
\approx G^{\mu\lambda }$, since our results are of $O(e)$ only.

It is worthwhile to make a comparison with Polyakov formulation at this stage.
The Lagrangian (69) is identical to the Polyakov one (43) in the conformal
gauge. There is a similar mapping between BCs (72) and (48) again
in the conformal gauge. Consequently, we
shall be reproducing the same set of modified brackets (34a) and (54), displaying noncommutativity among
various coordinates. It should be emphasised, however, that this agreement is only
upto ${\cal O}(e)$ in the coupling parameter in the specific gauge (62).

\section{Conclusion}
In this paper we have derived expressions for a noncommutative algebra
that are more general than the standard results found in the conformal gauge. Indeed,
 our results reproduce the standard ones, once the conformal gauge is implemented.

The origin of any modification in the usual Poisson algebra is the presence of boundary
conditions. This phenomenon is quite well known for a free scalar field subjected to
periodic boundary conditions. We showed that its exact analogue is the conformal gauge
fixed free string, where the boundary condition is of Neuman-type. This led to a
modification only in the $\{X^{\mu }(\s ),\Pi_{\nu }(\s' )\}$ algebra, where the usual
Dirac delta function got replaced by $\Delta_+ (\s ,\s' )$. A more general type
of boundary condition  occurs in the gauge independent formulation of a free Polyakov
string. Using certain algebraic consistency requirements, we showed that the boundary
conditions in the free theory naturally led to a noncommutative structure among
the coordinates. This non-commutativity  vanishes in the conformal gauge, as expected.

The same technique was adopted for the interacting string. A more
involved boundary condition led to a more general type of
noncommutativity than has been observed before. Contrary to the
standard conformal gauge expressions, this noncommutative algebra
survives at all points of the string and not just at the
boundaries. Furthermore, in contrast to the free theory, this
noncommutativity cannot be removed in any gauge. We have also
shown that, the noncommutativity does not affect the usual
diffeomorphism algebra among the gauge generators. In the
conformal gauge, our results reduce to the standard
noncommutativity found only at the string end points.

A perturbative analysis of the noncommutativity has also been performed in the interacting
Nambu-Goto string. Surprisingly, the conformal gauge result in the Polyakov formulation is
reproduced in the Nambu-Goto scheme in the lowest nontrivial order in the Neveu-Schwarz coupling,
in an orthonormal-like gauge. It would be interesting to see if there is an alternative gauge
condition in which the above equivalence can be shown exactly.


\newpage

\begin{thebibliography}{99}
\bibitem{sw}N.Seiberg and E.Witten, JHEP 09 032(1999).
\bibitem{dn}For a review, see M.R.Douglas and N.A.Nekrasov,
arXiv: HEP-TH/0108158; R.J.Szabo,  arXiv: HEP-TH/0109162.
\bibitem{chu}C.-S.Chu and P.-M.Ho, Nucl.Phys. B550 (1999) 151.
\bibitem{som}V.Schomerus, JHEP 06 (1999) 030.
\bibitem{di}P.A.M.Dirac, {\it Lectures on Quantum Mechanics} (Yeshiva
University Press, New York, 1964).
\bibitem{ard}F.Ardalan, H.Arfaei and M.M.Sheikh-Jabbari, JHEP 9902 
(1999) 016; W.T.Kim and J.J.Oh, 
Mod.Phys.Lett. A15 (2000) 1597; C.-S.Chu and P.-M.Ho, 
Nucl.Phys. B568 (2000) 447.
\bibitem{bra}N.R.F.Braga and C.F.L.Godinho, HEP-TH/0110297.
\bibitem{tez}M.M.Sheikh-Jabbari and A.Shirzad, Eur. Phys. Jour. C19 (2001) 383;
K-I. Tezuka, hep-th/0201171
\bibitem{hrt}A.J.Hanson, T.Regge and C.Teitelboim, {\it Constrained
Hamiltonian System}, Roma, Accademia Nazionale Dei  Lincei, (1976).
\bibitem{zab}M.Zabzine, JHEP 0010 (2000) 042.
\bibitem{and}M.A De Andrade, M.A. Santos and I.V.Vancea, JHEP 0106 (2001) 026.
\bibitem{holt}See for example J.W.van Holten, {\it Aspects of BRST Quantisation },
HEP-TH/0201124.
\bibitem{mk}M.Kaku, {\it Introduction to Superstring Theory}, Springer-Verlag, 1988.
\bibitem{pol}J.Polchinski, {\it String Theory}, Vol. I, Cambridge University Press, 1998.
\end{thebibliography}
\end{document}